\documentclass[11pt]{article}

\usepackage{amsmath}
\usepackage{amsfonts}
\usepackage{epsfig}
\usepackage{graphics}
\usepackage{url}

\usepackage{color}

\begin{document}

\title{Implementation of mixed-dimensional models for flow in
  fractured porous media}


\author{Eirik Keilegavlen, Alessio Fumagalli, Runar Berge, Ivar Stefansson}


\maketitle

\begin{abstract}
  Models that involve coupled dynamics in a mixed-dimensional geometry
  are of increasing interest in several applications.  Here, we
  describe the development of a simulation model for flow in fractured
  porous media, where the fractures and their intersections form a
  hierarchy of interacting subdomains.  We discuss the implementation
  of a simulation framework, with an emphasis on reuse of existing
  discretization tools for mono-dimensional problems.  The key
  ingredients are the representation of the mixed-dimensional geometry
  as a graph, which allows for convenient discretization and data
  storage, and a non-intrusive coupling of dimensions via boundary
  conditions and source terms.
  This approach is applicable for a wide class of
  mixed-dimensional problems.  We show simulation results for a flow
  problem in a three-dimensional fracture geometry, applying both
  finite volume and virtual finite element discretizations.
\end{abstract}

\section{Introduction}
\label{author_mini4:sec:2}
Simulation models for real-life applications commonly must represent
objects with high aspect ratios embedded in the domains.  This includes both objects of
co-dimension~1, exemplified by fractures in a porous rock, and
co-dimension 2 models such as reinforced concrete and root systems.
Although the embedded objects occupy a small part of the
simulation domain, they can have a decisive impact on the system
behavior, thus their representation in the simulation model is
critical.  The small object sizes, and in particular the high aspect
ratio, make an equi-dimensional representation computationally
prohibitively expensive.  The standard simulation technique has
therefore been to apply homogenization to arrive at an, ideally,
equivalent upscaled model.  In recent years, advances in computational
power, modeling approaches and numerical methods have made resolving
the objects more feasible. This calls for the development of new simulation tools for
mixed-dimensional problems that allow for a high degree of reuse of
software designed for mono-dimensional simulations.

Here, we explore the implementation of a simulation model based on a
newly developed modeling framework for flow and transport in fractured
porous media
\cite{Keilegavlen_mini25:Boon2016,Keilegavlen_mini25:Boon2017}. The
model considers fractures as manifolds of co-dimension 1 that are
embedded in the simulation domain, and further allows for
intersections of fractures as objects of co-dimension 2 and
3.
Central to our implementation
is the representation of the mixed-dimensional problem as a graph, where
the nodes represent mono-dimensional problems, and the edges represent
couplings between subdomains.  Discretization internal to each subdomain is
then a matter of iterating over the nodes of the graph, and apply standard
numerical methods by invoking, possibly legacy, mono-dimensional discretizations.
Discretization of mixed-dimensional dynamics, which is commonly not
handled by existing software, is associated with the edges of the graph.
Depending on how the interactions are modeled, the implementation of the subdomain couplings
comes down to treatment of boundary conditions and source terms, both
of which are standard in most numerical tools.

Focusing on locally conservative methods, which are prevailing in
porous media applications, we consider both virtual element and finite
volume approaches for flow, in addition to finite volume techniques
for transport.  The simulation model discussed is implemented in the
software framework PorePy~\cite{Keilegavlen_mini25:Keilegavlen2017a},
and is available at \url{www.github.com/pmgbergen/porepy}.

\section{Mixed-dimensional flow in fractured porous media}

Let us consider a $N$-dimensional domain $\Omega \subset \mathbb{R}^N$,
typically $N=2$ or $3$, with outer boundary $\partial\Omega$. $\Omega$
represents the porous medium, which is composed of a $N$-dimensional domain
$\Omega^N$ (the rock matrix) and lower-dimensional domains
$\Omega^{N-1}, \ldots, \Omega^0$, representing fractures and possibly objects
of lower dimensions such as fracture intersections and intersections of
fracture intersections. We assume that $\Omega^{d-1} \not\subset \Omega^d$ for
$d=0,\ldots, N$, and $\Omega = \cup_{d}\Omega^d$.  Let $\Gamma$ denote the set
of internal boundaries between subdomains of different dimension. In each
dimension, we consider the flow of a single-phase incompressible fluid, with
governing equations stated on mixed form as
\begin{equation}\label{keilegavlen_mini25:eq:01}
  \begin{aligned}
    \vec{u}^d +\vec{K}\nabla p^d &= 0
    &&\mathrm{in}\ \Omega^d, d>0\\
    \nabla\cdot \vec{u}^d - [\![\vec{u}^{d+1} \cdot \vec{n}^{d+1}]\!]&=f^d
    &&\mathrm{in}\ \Omega^d, d\leq N \\
    \vec{u}^{d}\cdot{\vec{n}^{d}} + \kappa (p^{d-1} - p^{d})&= 0
    &&\mathrm{on}\ \Gamma.
  \end{aligned}
\end{equation}
\noindent
The unknowns are the Darcy velocity $\vec{u}$ and the pressure $p$, and the
data are the permeability ${K}$, the effective normal permeability $\kappa$
between the subdomains, and $f$ is sources and sinks.  We define
$\vec{u}^{N+1}=0$.  The coupling between pairs of subdomains one dimension
apart can be seen in the conservation statement, where the divergence of the
flux in the dimension under consideration is balanced by source terms in the
same dimension, and the contribution from the higher dimension via the jump
$[\![\vec{u}^{d+1} \cdot \vec{n}^{d+1}]\!]$, which represent the net flux
into the domain. From the higher dimension, this
term will appear as the leakage into lower dimensions, which may be interpreted
as an internal boundary.  The last equation in \eqref{keilegavlen_mini25:eq:01}
is a mixed-dimensional Darcy law with effective permeability $\kappa$,
which models the flux between two subdomains
separated by the interface $\Gamma$. On outer boundaries of each $\Omega^d$
we assign Dirichlet or Neumann conditions.  For more details on the
mathematical model, we refer to
\cite{Keilegavlen_mini25:Martin2005,Keilegavlen_mini25:DAngelo2011,Keilegavlen_mini25:Fumagalli2012g,Keilegavlen_mini25:Schwenck2015,Keilegavlen_mini25:Brenner2016a,Keilegavlen_mini25:Boon2016,Keilegavlen_mini25:Scotti2017,Keilegavlen_mini25:Chave2017}.
We also note that a similar model can be used to express transport of a scalar
in mixed-dimensional geometries, see \cite{Keilegavlen_mini25:Fumagalli2017d}.

\section{Discretization of mixed-dimensional problems}

To device design principles for an implementation of the model
\eqref{keilegavlen_mini25:eq:01}, we first observe that interaction between
subdomains takes the form of boundary conditions from lower to higher
dimension and source terms.
Moreover, this should apply also for more general classes of models, including
most, if not all, that are built upon conservation principles
\cite{Keilegavlen_mini25:Boon2017}.  Thus a versatile implementation should be
based on independent discretizations on the subdomains, together with
appropriate coupling conditions.  Below we describe how this is naturally
achieved by representing the computational grid as a graph, and discretization
as an iteration over its nodes and edges.  This abstraction allows for reuse of
an existing code base, and if combined with a flexible interface between the
grids, it can be extended to heterogeneous discretizations and multi-physics
modeling.

In Subsection \ref{subsec:grid} we present the grid structure needed for the
mixed-dimensional description of \eqref{keilegavlen_mini25:eq:01}, while in
Subsection \ref{subsec:discr} we briefly introduce the numerical
discretizations.

\begin{figure}
  \centering \includegraphics[width=.75\textwidth]{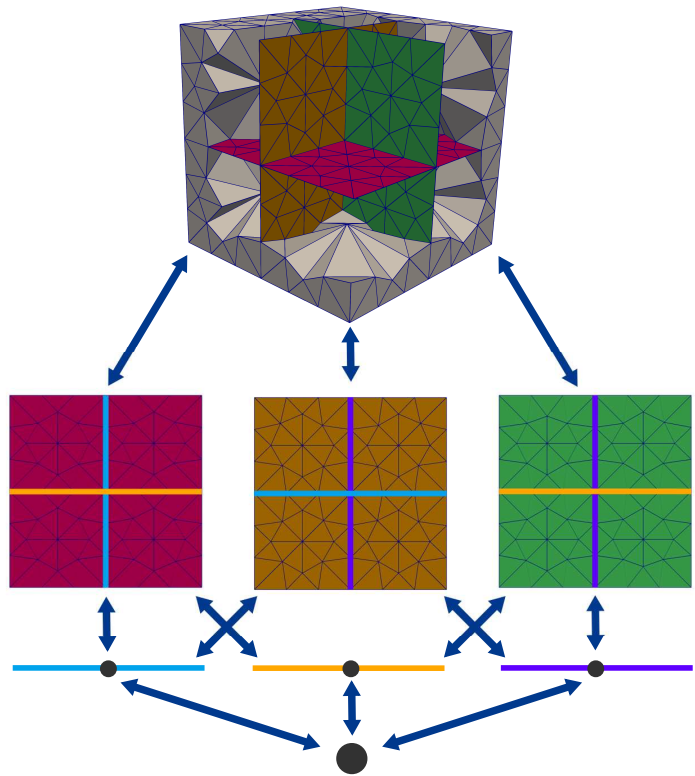}
  \caption{Hierarchy of grids derived from the meeting of three 2D objects
    embedded in 3D space.  The mixed-dimensional grid is naturally represented
    as a graph, with individual grids forming nodes.}
  \label{keilegavlen_mini_25_fig:1}
\end{figure}
\begin{figure}
  \def\svgwidth{0.4\textwidth} \centering \begingroup%
  \makeatletter%
  \providecommand\color[2][]{%
    \errmessage{(Inkscape) Color is used for the text in Inkscape, but the
      package 'color.sty' is not loaded}%
    \renewcommand\color[2][]{}%
  }%
  \providecommand\transparent[1]{%
    \errmessage{(Inkscape) Transparency is used (non-zero) for the text in
      Inkscape, but the package 'transparent.sty' is not loaded}%
    \renewcommand\transparent[1]{}%
  }%
  \providecommand\rotatebox[2]{#2}%
  \ifx\svgwidth\undefined%
  \setlength{\unitlength}{140.79999188bp}%
  \ifx\svgscale\undefined%
  \relax%
  \else%
  \setlength{\unitlength}{\unitlength * \real{\svgscale}}%
  \fi%
  \else%
  \setlength{\unitlength}{\svgwidth}%
  \fi%
  \global\let\svgwidth\undefined%
  \global\let\svgscale\undefined%
  \makeatother%
  \begin{picture}(1,1.00001283)%
    \put(0,0){\includegraphics[width=\unitlength]{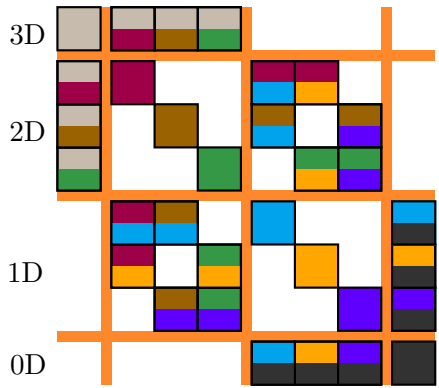}}%
    \put(-0.02697299,0.90591653){\color[rgb]{0,0,0}\makebox(0,0)[rb]{\smash{3D}}}%
    \put(-0.02647361,0.65166972){\color[rgb]{0,0,0}\makebox(0,0)[rb]{\smash{2D}}}%
    \put(-0.03271587,0.2809165){\color[rgb]{0,0,0}\makebox(0,0)[rb]{\smash{1D}}}%
    \put(-0.02522517,0.03814863){\color[rgb]{0,0,0}\makebox(0,0)[rb]{\smash{0D}}}%
    \put(-0.02556823,0.85512644){\color[rgb]{0,0,0}\makebox(0,0)[lt]{\begin{minipage}{0\unitlength}\raggedright
        \end{minipage}}}%
  \end{picture}%
  \endgroup%
  \caption{Structure of linear system resulting from a discretization of the
    mixed-dimensional problem on the geometry shown in
    Figure~\ref{keilegavlen_mini_25_fig:1}. The diagonal blocks represent the
    discretization of each grid, while the off diagonal blocks represent the
    interface condition between two and two grids. The top and bottom colors of
    each coupling block correspond to the higher and lower dimensional grids,
    respectively.}
  \label{keilegavlen_mini_25_fig:2}
\end{figure}

\subsection{Grid structure}\label{subsec:grid}
For simplicity, we require that the computational mesh is fully conforming to
all objects $\Omega^d$ in all dimensions. This condition can be relaxed,
e.g. by applying mortar discretizations in the transition between subdomains,
however, we will not pursue this herein.

To illustrate the mixed-dimensional grid structure, we consider the hierarchy
of grids depicted in Figure 1. The main domain is $\Omega = [-1, 1]^3$. Define
the fracture $\Omega_1^2 = [-1,1] \times [-1, 1] \times \{0\}$, and similarly
let $\Omega_2^2$ and $\Omega_3^2$ be embedded in the $xz$ and $yz$ plane,
respectively. The intersection between pairs of fractures defines intersection
lines $\Omega_i^1, i=\{1, 2, 3\}$ along the coordinate axes. Further, the
intersection lines intersect to define $\Omega^0$ in the origin.  Finally,
define
$\Omega^3 = \Omega \setminus \big(\bigcup_{d={0,1,2}}\bigcup_{i}
\Omega_i^d\big)$.

To apply a discretization scheme on this grid structure requires iterations
over all subdomains and their connections. This is readily implemented by
considering the subdomains as nodes in a graph, with the connections forming
edges, as illustrated in Figure 1. We define $nodes(\cdot)$ as a function such
that given a dimension $d$ it returns the nodes in the graph of the same
dimension. Similarly, $edges(\cdot,\cdot)$ is a function that, given
consecutive dimensions $d$ and $d-1$, returns the edges in the graph associated
with nodes of dimension $d$ and $d-1$.  This abstraction is particularly
suitable for existing software frameworks that can handle grids of a single but
flexible dimension. Discretization of a mixed-dimensional problem can then be
defined by two iterations; (i) an iteration on the nodes of the graph to invoke
the standard solver on each subdomain, and (ii) an iteration on the edges of
the graph to impose coupling conditions on subdomain boundaries.

Given a suitable numerical scheme for the discretization of
\eqref{keilegavlen_mini25:eq:01}, the system obtained from the first iteration
gives the following block-diagonal matrix
\begin{gather*}
  {\rm diag}(A^N, A^{N-1}, \ldots, A^1, 0),
\end{gather*}
where $A^d$ is related to the set $nodes(d)$ of nodes in the graph. The
matrices $A^d$ are block diagonal, with one block for each node defining the
mono-dimensional discretization of each sub-grid.  The interface conditions
between $d$ and $d-1$ can be written as
\begin{gather*}
  \begin{bmatrix}
    H^{d, d-1}    & C^{d, d-1}\\
    C^{d, d-1} & L^{d, d-1}
  \end{bmatrix}
\end{gather*}
where the matrices $C^{d,d}$ and $C^{d-1,d-1}$ are contributions of the
coupling condition for the higher and lower dimension, both having a block
diagonal structure.  The matrix $C^{d, d-1}$ represents the inter-dimensional
contribution and has a block structure.  All three matrices contain one block
for each edge in $edges(d, d-1)$. For a three-dimensional problem, the general
structure of the global matrix is
\begin{gather*}
  \begin{bmatrix}
    A^3+H^{3, 2} & C^{3,2}     & 0           & 0        \\
    C^{3,2}      & A^2+L^{3,2}+H^{2,1} & C^{2,1}     & 0        \\
    0            & C^{2,1}     & A^1+L^{2,1}+H^{1,0} & C^{1,0}  \\
    0            & 0           & C^{1,0}     & L^{1,0}        \\
  \end{bmatrix}.
\end{gather*}
The matrix in Figure \ref{keilegavlen_mini_25_fig:2} illustrates the block
structure associated with the mesh of Figure \ref{keilegavlen_mini_25_fig:1}.

\subsection{Conservative discretizations}\label{subsec:discr}
We consider two discretization schemes for the mixed-dimensional pressure
equation \eqref{keilegavlen_mini25:eq:01}. The simplest option is a finite
volume scheme built as a two-point flux approximation (TPFA), which is standard
in commercial porous media simulators. This scheme is easy to implement, and,
with the data structures outlined above, a simple extension to
mixed-dimensional problems is fairly straightforward; more complex approaches
involving mortar variables are currently under investigation.
However, TPFA is consistent only for K-orthogonal grids, and can be expected to
suffer from poor accuracy for the complex grids needed to cover realistic
fracture networks.  Our second approach applies the dual form of the virtual
element method \cite{Keilegavlen_mini25:BeiraoVeiga2016}, with the extension to
mixed-dimensional problems, as discussed in
\cite{Keilegavlen_mini25:Fumagalli2017d}.  The virtual element method puts
almost no restrictions on the cell shape, and is thus ideally suited for
handling rough grids.  This also makes it possible to merge simplex cells into
general polyhedral shapes, and thus reduce the number of degrees of freedom. We
do not consider non-simplex cells here, see
\cite{Keilegavlen_mini25:Fumagalli2017d} for details.

Given a flux field, the extension of the tracer advection problem using a
first-order upwind scheme is equivalent to the procedure above.

\section{Example simulation}
In this part we present an example to assess the above models and
numerical schemes. The source code of the example is available online
in the PorePy repository.
\begin{figure}[!t]
  \centering
  \includegraphics[width=0.45\textwidth]{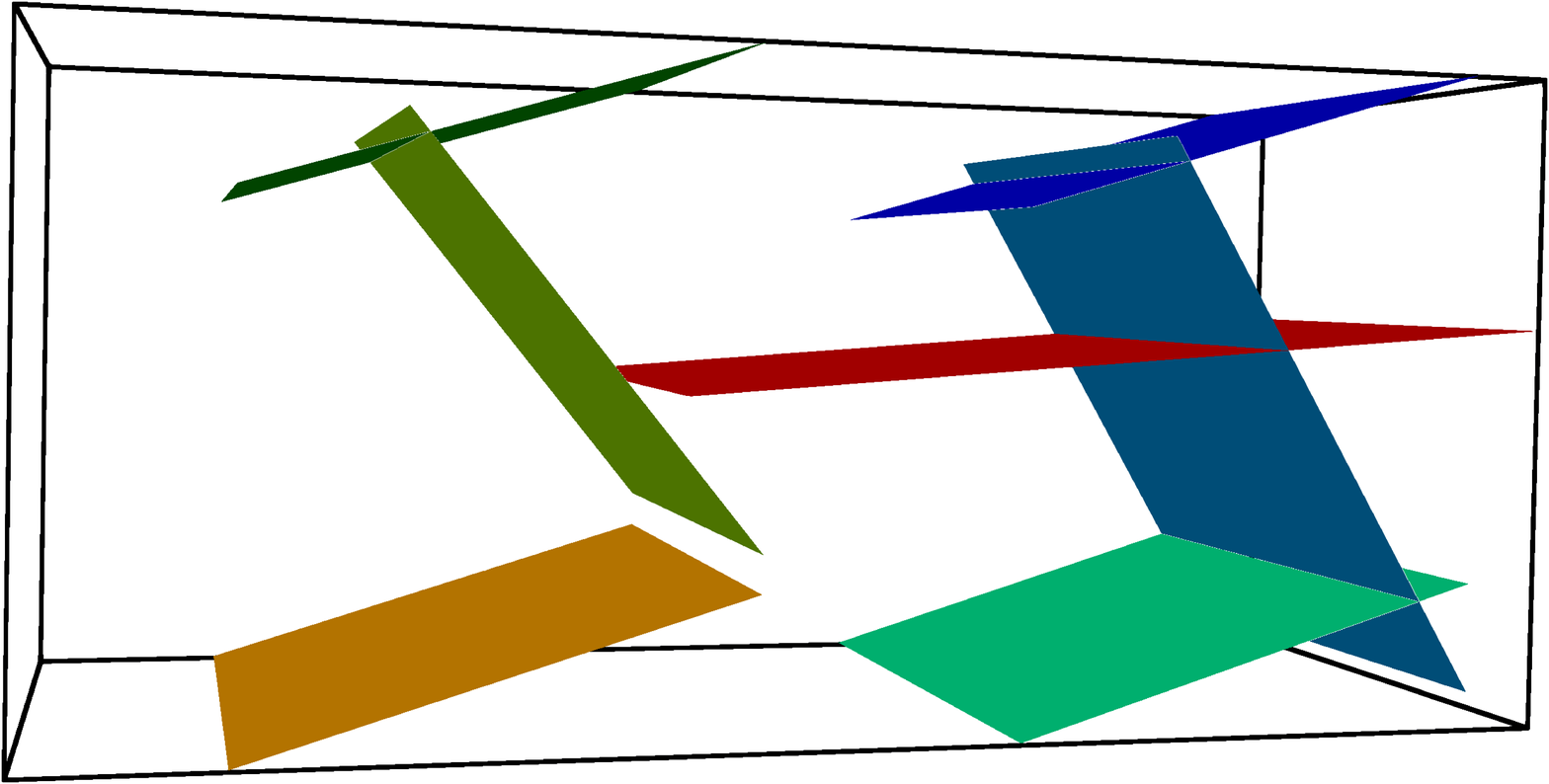}\\
  \includegraphics[width=0.45\textwidth]{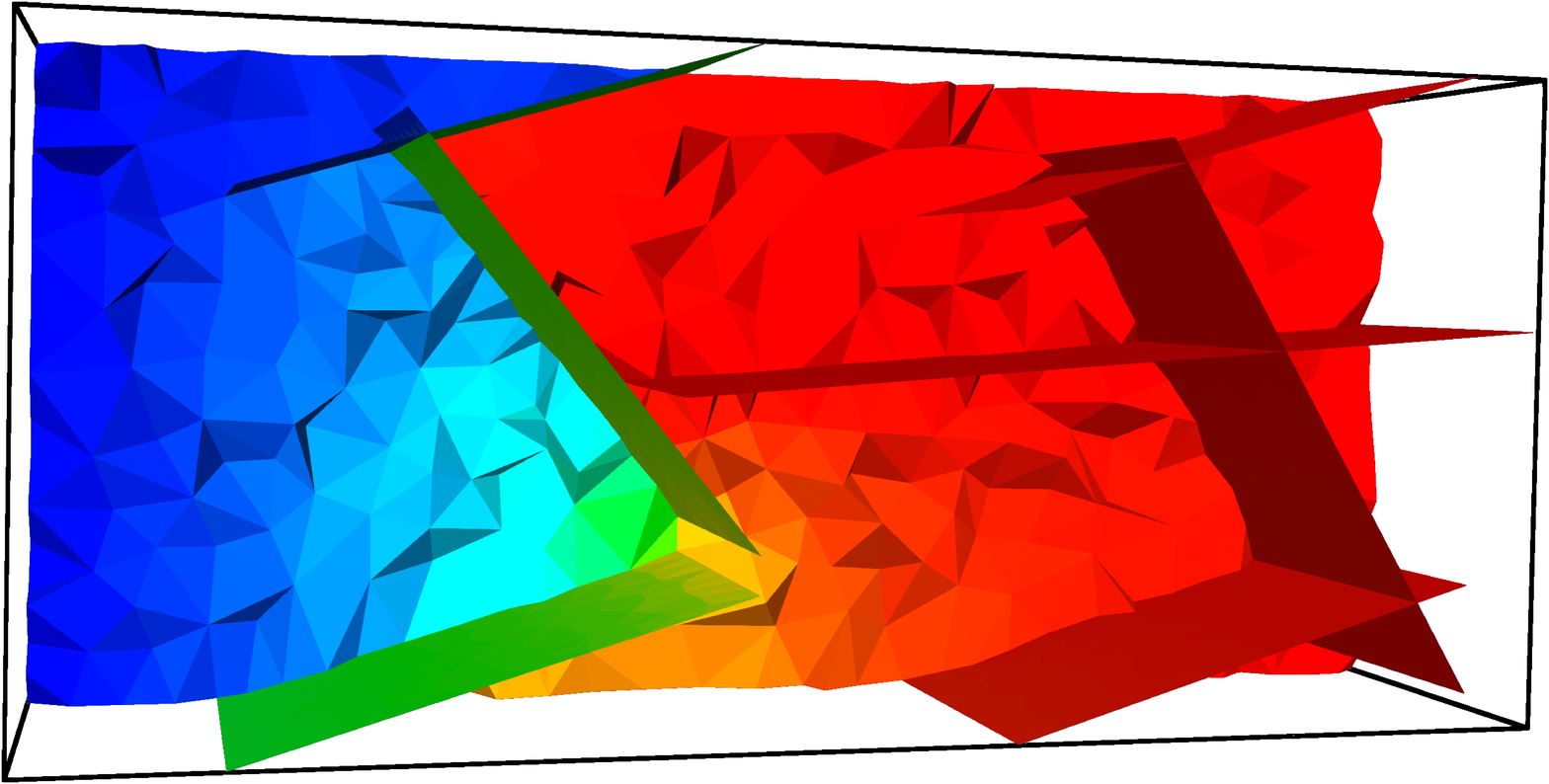}\hfill%
  \includegraphics[width=0.45\textwidth]{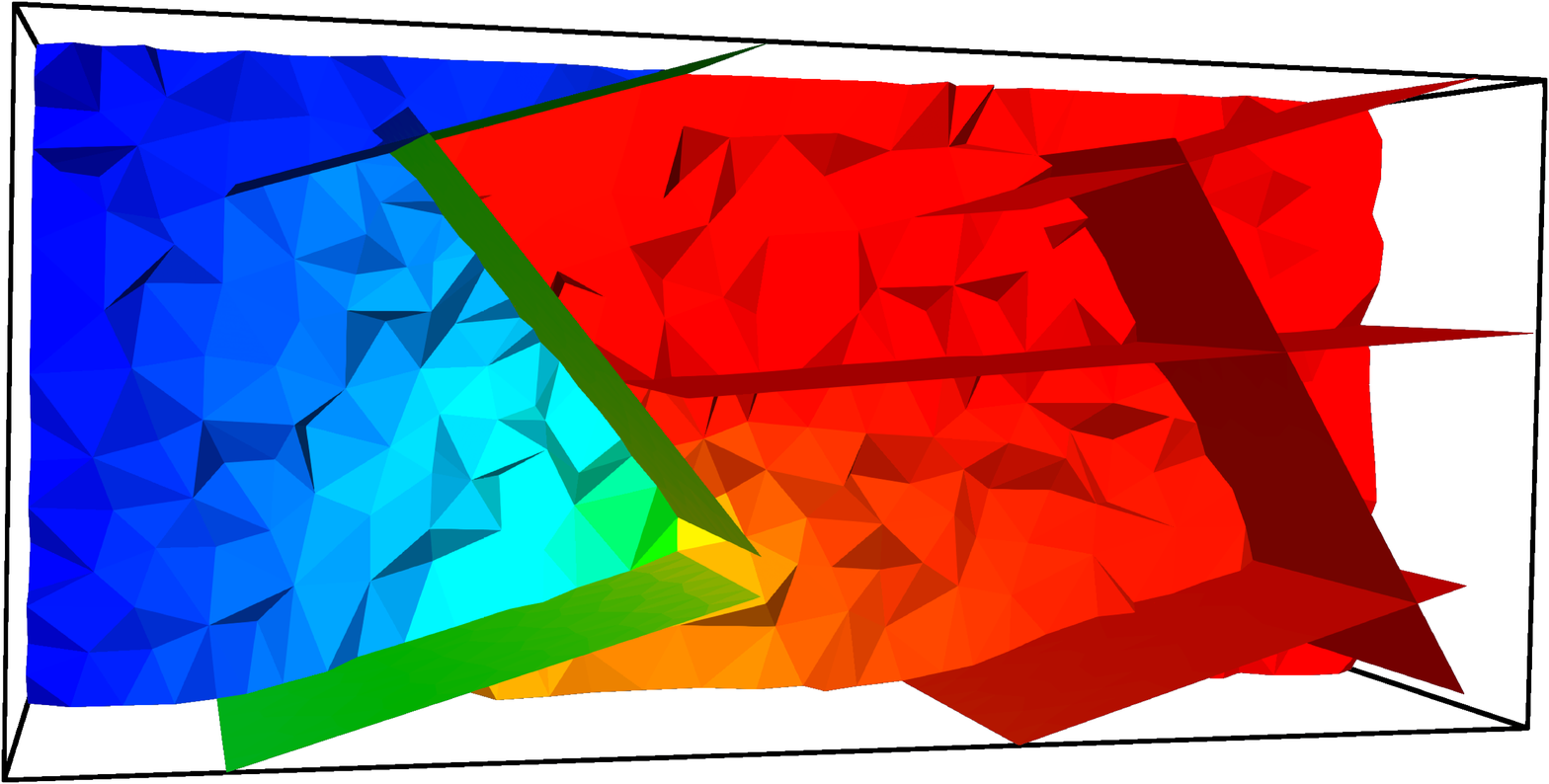}\\
  \includegraphics[width=0.45\textwidth]{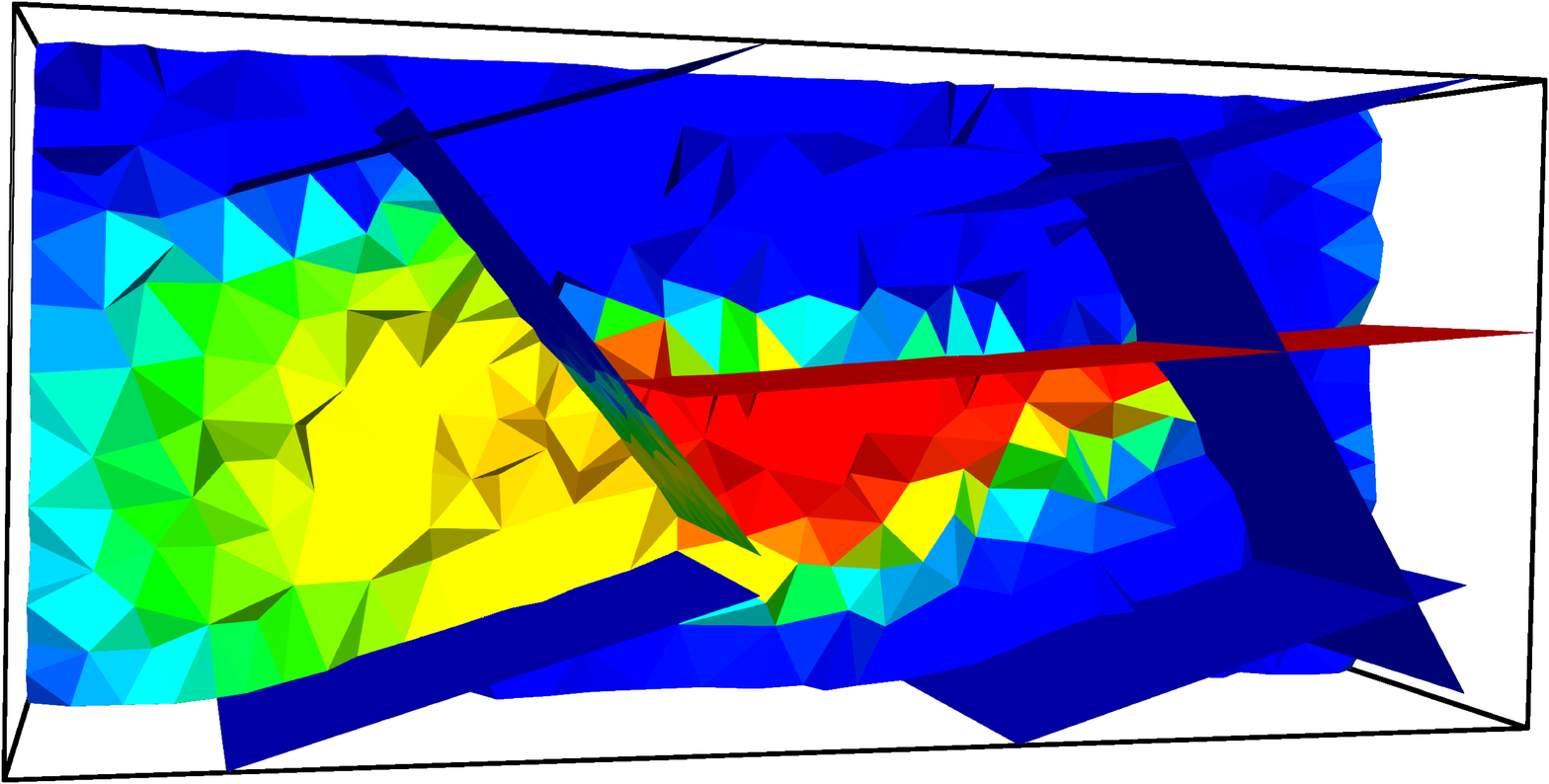}\hfill%
  \includegraphics[width=0.45\textwidth]{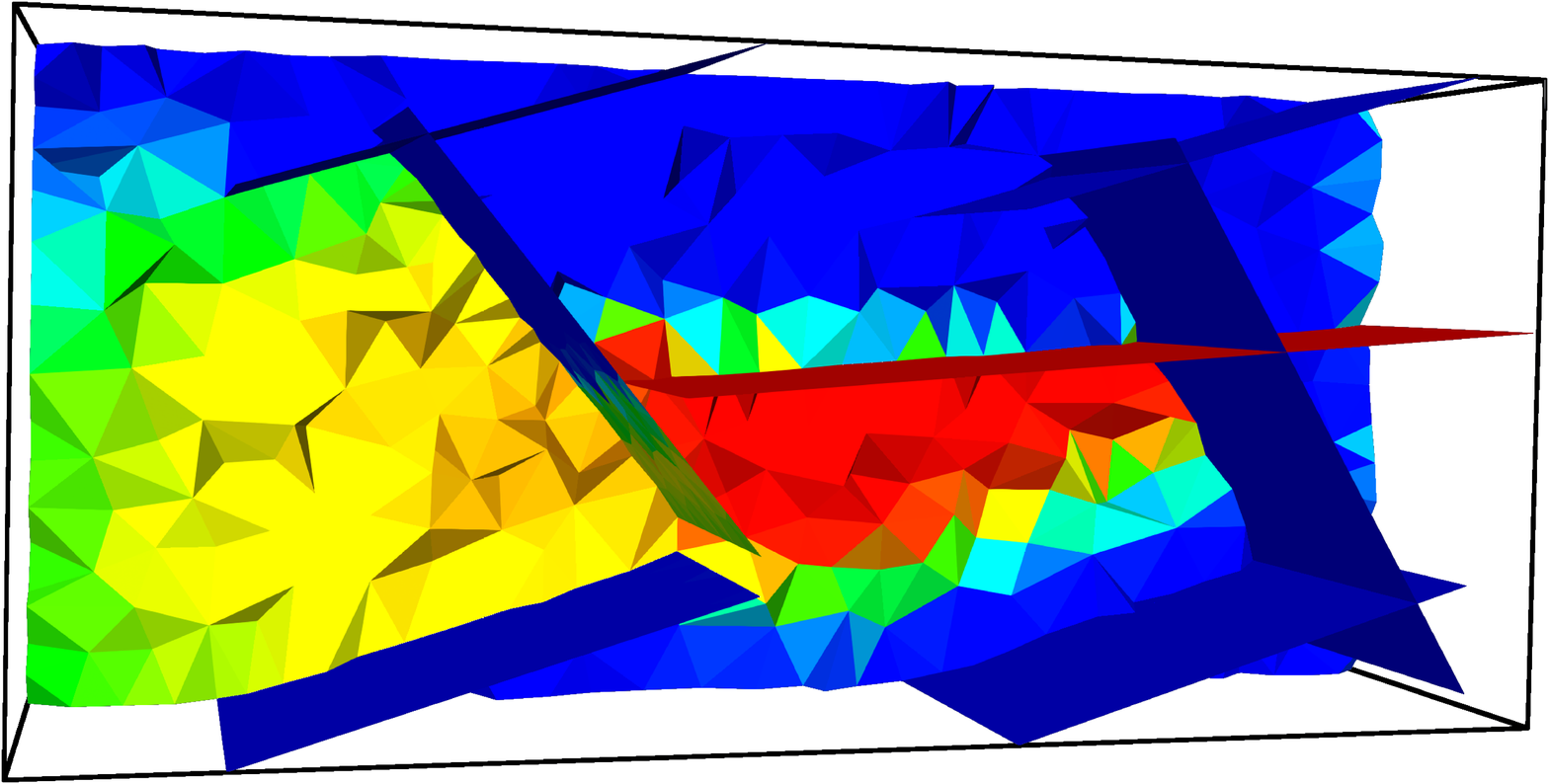}
  \caption{Top: representation of the considered geometry for the
    example.  Centre: pressure field for the TPFA (left) and VEM
    (right). Bottom: concentration at the final time with the
    discharge computed by the TPFA and VEM. The flow is from the right
    to the left.  In all the figures a ``Blue to Red Rainbow'' colour
    map is used in the range $[0,1]$.}%
  \label{fig:sol}
\end{figure}
We consider a 3D fracture network with heterogeneous permeabilities in
the fractures and on their intersections exploiting the
mixed-dimensional structure of the model. The domain is
$\Omega= [0, 2]\times[0, 1]\times[0, 1]$ and the geometry of the
network, containing 7 fractures, is depicted at the top of Figure
\ref{fig:sol}. The permeability in the rock matrix is assumed
unitary. The fracture marked with a red color in the figure is high
permeable with permeability equal to $10^4$, while the other fractures
behave as barriers having permeabilities equal to $10^{-4}$. The
fracture aperture is constant and equal to $10^{-2}$ for all
fractures.
All 1D intersections inherit the highest permeability of the
intersecting fractures. Thus, the ones involving the highly permeable
fracture are permeable, and the others behave as impermeable
paths. The grid is composed by 23325 tetrahedra for the 3d rock
matrix, 2624 triangles for the fractures, and 48 segments for the
fracture intersections.

We impose a pressure gradient by setting the pressure on the left
boundary to 0 and the right boundary to 1. The other boundaries are
given a zero flux condition.  We compute the pressure field and the
discharge (flux) using the TPFA and VEM on the same grid. The
discharge is a face variable, obtained by back-calculation from the
pressures in the TPFA method and directly computed for the VEM.  The
computed solution is represented in the middle row of Figure
\ref{fig:sol}. The low permeable fractures at the right end of the
domain do not affect the pressure field much, as the conductive fracture
makes a high permeable connection between the right and the center
part of the domain. However, at the left part the low permeable
fractures force a pressure gradient between the small gap between two
of the fractures (orange and green). The solutions obtained from both
methods are in good agreement.

Once the discharge is computed we consider a pure transport problem,
where the advective field is given by the discharge. We inject a
tracer with concentration 1 on the right part of the domain with
outflow on the left part. The tracer thus follows the discharge, flowing
in the high permeable fracture first and then propagating in the rock
matrix, avoiding the low permeable fractures in the left part of the
domain. An implicit Euler scheme is applied for the time
discretization with time step equal to $0.01$. The final time of the
simulation is $3$. As in the case of the pressures, the tracer
solutions for the discharge computed by the TPFA or by the VEM are in
agreement.

\section{Concluding remarks}
We have discussed the design of simulation tools for mixed-dimensional
equations, in the setting of flow in fractured porous media.  We
presented a data hierarchy where the mixed-dimensional structure is a
graph, with each node representing a standard mono-dimensional domain.
This allows for extensive reuse of existing code designed for
mono-dimensional problems, and also facilitates simple implementation
of new discretization schemes.  Numerical examples of flow and
transport in a three-dimensional fractured medium illustrate the
capabilities of a simulation tool based on this approach.

\section*{Acknowledgments}

We acknowledge financial support from the
Research Council of Norway, project no.  244129/E20 and 250223.

%
%
%


\begin{thebibliography}{99}
  \parskip1.0ex

\bibitem{Keilegavlen_mini25:Boon2017} {\sc W.~M. Boon,
    J.~M. Nordbotten, and J.~E. Vatne}, {\em Mixed-dimensional
    elliptic partial differential equations}, arXiv:1710.00556, 2017.

\bibitem{Keilegavlen_mini25:Boon2016} {\sc W.~M. Boon,
    J.~M. Nordbotten, and I.~Yotov}, {\em Robust discretization of
    flow in fractured porous media}, Tech. report, arXiv:1601.06977v2,
  2017.

\bibitem{Keilegavlen_mini25:Brenner2016a}
{\sc K.~Brenner, J.~Hennicker, R.~Masson, and P.~Samier}, {\em {Gradient
  discretization of hybrid-dimensional Darcy flow in fractured porous media
  with discontinuous pressures at matrix-fracture interfaces}}, {IMA Journal of
  Numerical Analysis} (2016).


\bibitem{Keilegavlen_mini25:Chave2017}
{\sc F.~A. Chave, D.~Di~Pietro, and L.~Formaggia}, {\em {A Hybrid High-Order
  method for Darcy flows in fractured porous media}}, Tech. report, HAL
  archives, 2017.

\bibitem{Keilegavlen_mini25:DAngelo2011} {\sc C.~D'Angelo and
    A.~Scotti}, {\em A mixed finite element method for {D}arcy flow in
    fractured porous media with non-matching grids}, Mathematical
  {M}odelling and {N}umerical {A}nalysis {\bf 46}:02 (2012), 465--489.

\bibitem{Keilegavlen_mini25:Fumagalli2017d} {\sc A.~Fumagalli and
    E.~Keilegavlen}, {\em Dual virtual element methods for discrete
    fracture matrix models}, arXiv:1711.01818, 2017.

\bibitem{Keilegavlen_mini25:Fumagalli2012g} {\sc A.~Fumagalli and
    A.~Scotti}, {\em An {E}fficient {XFEM} {A}pproximation of {D}arcy
    {F}lows in {A}rbitrarily {F}ractured {P}orous {M}edia}, {O}il and
  {G}as {S}ciences and {T}echnologies - {R}evue d'{IFP} {E}nergies
  {N}ouvelles {\bf 69}:4 (2014), 555--564.

\bibitem{Keilegavlen_mini25:Keilegavlen2017a} {\sc E.~Keilegavlen,
    A.~Fumagalli, R.~Berge, I.~Stefansson, and I.~Berre}, {\em Porepy:
    An open source simulation tool for flow and transport in
    deformable fractured rocks}, arXiv:1712.00460, 2017, In
  preparation for Computer \& Geosciences.

\bibitem{Keilegavlen_mini25:Martin2005} {\sc V.~Martin, J.~Jaffr{\'e},
    and J.~E. Roberts}, {\em Modeling {F}ractures and {B}arriers as
    {I}nterfaces for {F}low in {P}orous {M}edia}, SIAM J. Sci.
  Comput. {\bf 26}:5 (2005), 1667--1691.

\bibitem{Keilegavlen_mini25:BeiraoVeiga2016} {\sc L.~Beir\~{a}o~da
    Veiga, F.~Brezzi, L.~D. Marini, and A.~Russo}, {\em Mixed virtual
    element methods for general second order elliptic problems on
    polygonal meshes}, ESAIM: M2AN {\bf 50}:3 (2016), 727--747.

\bibitem{Keilegavlen_mini25:Schwenck2015}
{\sc N.~Schwenck, B.~Flemisch, R.~Helmig, and B.~Wohlmuth}, {\em Dimensionally
  reduced flow models in fractured porous media: crossings and boundaries},
  Computational Geosciences {\bf 19}:6 (2015), 1219--1230 (English).

\bibitem{Keilegavlen_mini25:Scotti2017}
{\sc A.~Scotti, L.~Formaggia, and F.~Sottocasa}, {\em Analysis of a mimetic
  finite difference approximation of flows in fractured porous media}, ESAIM:
  M2AN (2017).

\end{thebibliography}

\end{document}